\documentclass[conference]{IEEEtran}
\IEEEoverridecommandlockouts
\usepackage{amsmath,amssymb,amsfonts}
\usepackage{algorithmic}
\usepackage{graphicx}
\usepackage{textcomp}
\usepackage[dvipsnames]{xcolor}
\usepackage[style=numeric-comp,sorting=none,sortcites]{biblatex}
\newcommand{\coo}{\ensuremath{\mathrm{CO_2}}}
\begin{document}
\title{Green Resilience of Cyber-Physical Systems\\
{}
\thanks{}
}
\author{\IEEEauthorblockN{Diaeddin Rimawi}
\IEEEauthorblockA{\textit{Faculty of Science and Technology} \\
\textit{Free University of Bozen-Bolzano}\\
Bolzano, Italy \\
drimawi@unibz.it}}
\maketitle
\begin{abstract}
Cyber-Physical System (CPS) represents systems that join both hardware and software components to perform real-time services. Maintaining the system's reliability is critical to the continuous delivery of these services. However, the CPS running environment is full of uncertainties and can easily lead to performance degradation. As a result, the need for a recovery technique is highly needed to achieve resilience in the system, with keeping in mind that this technique should be as green as possible. This early doctorate proposal, suggests a game theory solution to achieve resilience and green in CPS. Game theory has been known for its fast performance in decision-making, helping the system to choose what maximizes its payoffs. The proposed game model is described over a real-life collaborative artificial intelligence system (CAIS), that involves robots with humans to achieve a common goal. It shows how the expected results of the system will achieve the resilience of CAIS with minimized \coo{} footprint.
\end{abstract}
\section{Introduction}
A Cyber-Physical System (CPS) is known for its heterogeneous components and the collaboration between its physical parts and the software controllers to perform real-time services. Studies that handle CPS are dramatically increasing. Each study handles one or more non-functional property of the system, like security \cite{liu_dynamic_2017},
performance \cite{rosales_actor-oriented_2018}, 
reliability \cite{oumimoun_solo-checkpointing_2021}, and resilience \cite{januario_distributed_2019}
to mention a few.
The systems complexity varies from one system to the other, like decoders \cite{liu_dynamic_2017}, power systems \cite{wu_resilience-based_2021}, smart vehicles \cite{akowuah_recovery-by-learning_2021}, smart spaceflight \cite{iv_feasibility_2018}, to smart robots that interact with the human \cite{pagliari_case_2017}, and list goes on.

Working robots especially collaborative artificial intelligence systems (CAIS) are one of the CPS types, where the human and robot are working together to achieve a common goal \cite{camilli_risk-driven_2021}. In these systems, online learning done by the robot through monitoring the human is normally combined to help the human achieve the goal.
The specialty of this type of CPSs, specifically human involvement, raises many challenges to study. The risk of working in the same physical space, the need to reduce human interaction as much as possible, and reduce wrong and unexpected movement of the robot, and other more, are all challenges to be addressed in CAIS.

Game theory is the study of two or more adversaries who play to choose the best action from their perspective to maximize their payoffs, through a reliable solution model for fast decision-making \cite{mei_engineering_2017, flokas_no-regret_2020}.

This early doctorate proposal studies the system reliability by discussing the system resilience and ability to recover performance degradation caused by uncertainty in the online learning and classification process. Moreover, it proposes a green resilience by considering minimizing the \coo{} footprints while recovering the system. 
The research proposes a game theory based solution, that considers green and resilience as adversaries where they can make decisions based on their own interests.

The rest of this doctorate proposal includes state of art done until now for green and resilience in CPS, then a description of the technical problem and the proposed solution using game theory is in the proposal section. The conclusion with future work and the doctorate timeline are discussed in the last section.

\section{State of the art}

\footnote{All references from the state-of-the-art have been omitted for space reasons, please refer to the following repository for the full version of the references ``\url{https://bit.ly/3wvJXjv}"}This work defines a cyber-physical system (CPS) as an ``employed system to provide complex real-time services by controlling the system physical parts through the system computational algorithms based on the monitored environment input".

\textbf{Resilience}: Resilience has various definitions based on its fields, like psychology, psychiatry, computer science, and other fields.
This proposal focuses on resilience for CPS. The fast ability to detect and recover from performance faults, is one definition, 
another one describes resilience as the system's ability to stand against extreme disasters. Moreover,
 another study has described two properties of CPS resilience, \textit{internal autogenous resilience} that refers to the system's ability to detect and process faults and attacks, and \textit{external resilience}, which is the maintenance of a safe operation within the system surrounding environment.

This proposal discusses the internal resilience of CPS, and it refers to resilience as ``the ability of the system to detect performance degradation and perform one or more actions to first mitigate the degradation and finally return to the original state in the fastest time possible".

Several studies focus on achieving resilience in CPS different components of the system, like using a hierarchical multi-agent framework to insure an accepted performance rate in case of any physical disturbance and cyber attacks. On the other hand, a tri-optimization model was used to identify the threat's capabilities to handle resilience against malicious threats in the power grid CPS. Another research has addressed resilience in a food industrial CPS by modeling each component as a smart machine equipped with a set of recovery services, through sensor Data API that collects data acquired from the physical side to monitor its behavior, and an Operator to detect abnormal conditions and pushing recovery actions to on-field operators.
\enlargethispage{\baselineskip}

Hierarchical-based solutions were proposed to enhance resilience by having a distributed resilience manager that utilizes the concept of management hierarchy, which ensures faster fault recovery. In the same context, another research has proposed an automated framework to guarantee resilience in CPS. The framework is based on a hierarchical contract-based, where it monitors the system components' non-functional properties through Assume Guarantee (A-G) contracts, and it refines these contracts to lower-level ones based on the I/O dependencies between the system components. This structure allows the framework to identify the root programmatic contract, then apply a multi-objective optimization problem to search for the optimal parameters of each lower-level contract.

Most of the approaches consider the physical part of the system and depend on the sensor's input to determine the system status to study the resilience of CPS facing malicious attacks like data injection.

A recent survey discussing CPS resilience has exposed several research gaps that need more investigation. The survey shows that there is a lack of research done on systems that involve humans to improve CPS resilience (e.g. CAIS), and shows the need for more studies that consider quantitatively CPS modeling.

\textbf{Green}: Information technological systems (IT-systems) are responsible for storing, displaying, transforming, and transferring information, and CPS is one of these systems. Green is defined in these systems as ``the efficient usage of energy with minimizing adverse effects", and \coo{} emission footprint is one of the known direct and indirect negative effects of CPS, during development and application. To measure \coo{} footprint, CodeCarbon is a Python lightweight package that estimates the \coo{} footprint coming from cloud/local computing resources. It can wrap the overall computation code and gives the estimation based on the geographical location of the system under testing.

More work has been made to discuss green in CPS and its effect on resilience. A recent study has studied the effect of green infrastructure (GI) location on sewer system resilience. However, their results were to motivate the placement importance of the GI in the sewer network. Other works were focusing on the effect of green strategies on the business/financial growth and resilience of the organization, talking about the greenhouse gas consumption for smart cities, and the effect on the cities' resilience, or discussing the manufacturing material for designing components that supports CPS, although these studies do not discuss the effect of computational processes on energy consumption. 
\enlargethispage{\baselineskip}
A new three-objectives optimization solution to address the number of facilities needed for a meat supply chain, to maintain the system economics, green, and resilience (eco-greslient), by optimizing the output from a fuzzy AHP (analytical hierarchy process) that determines the resilience pillars (robustness, agility, leanness, and flexibility) weights has been presented by recent research.
Another work has addressed the pattern that causes Google Tensor Processing Unit (TPU) error of activating sequences in the systolic array, which leads to a minimal loss in prediction accuracy, however, low-voltage operation, which is critical to reducing the energy adverse like \coo{}.

This research proposes the use of CodeCarbon \cite{mila_codecarbonio_nodate} to give a quantitative status of the computational \coo{} footprint to monitor the Green Resilience CPS.
\enlargethispage{\baselineskip}
\section{Proposal}
\textbf{Case Study}:
This proposal adopts a real-life CPS case that involves a robotics arm responsible for categorizing objects based on their colors and placing them in the corresponding boxes. The arm learns the corresponding box of each color during an online learning process, where a human operator picks the objects and places them in their boxes while the robot monitors through a visual machine learning classifier. This special type of CPS is called Collaborative AI systems (CAIS), which are robotic systems that collaborate with humans in a shared physical space to achieve common goals \cite{camilli_risk-driven_2021}.
The classifier is monitoring the object to be classified by subscribing to an RGB camera installed above the conveyor belt (conveyor hereinafter) and streams the object image. Secondly, it monitors the human operator's movements by subscribing to a Kinect camera that detects the human skeleton and streams it. The classifier learns with structured heterogeneous data sources associated with features (object color, conveyor speed, and object spatial) to yield category labels as an output.

\begin{figure}[t]
\centerline{\includegraphics[width = 0.38 \textwidth]{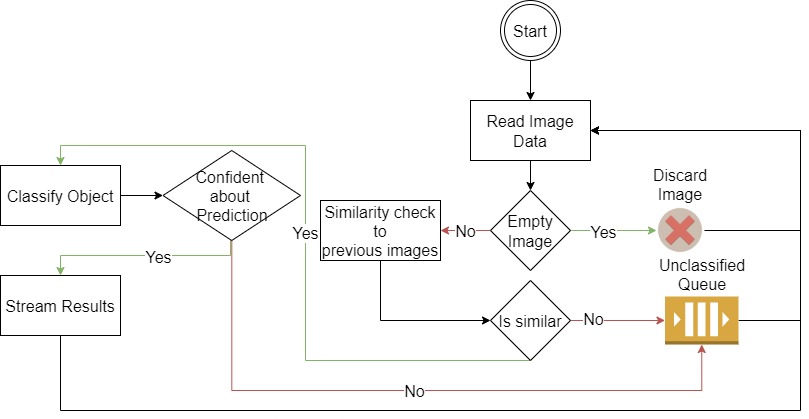}}
\caption{Online Classification Workflow.}
\label{online_workflow}
\end{figure}

Fig.~\ref{online_workflow} shows the online classification workflow. The online classification starts when the RGB camera streams new image data. The streamed image might be empty, so the robot arm discards the image and stays idle. If the image is not empty, the classifier runs a similarity check with previous images data, in case the image is not similar to previous images, it sends the image to the unclassified queue (learning mode, the human operator either classifies it or discards it). Then it classifies the object image and sends the image to the unclassified queue in case it is not confident about the prediction, while if it is confident it streams the results to the robot arm controller to move and place the object in the corresponding box, based on the previous learning. The confidence level of the classification is a quantitative estimation of the prediction made by the classifier \cite{poggi_quantitative_2017}.
\enlargethispage{\baselineskip}
\enlargethispage{\baselineskip}
In the case of misclassification by the robot arm, the human can validate the robot's action and correct the classifier. Miss classifying the object's color can be a result of many uncertain circumstances, like (I) the object color is still not classified by the human operator. This is due to the online learning process and the unknown set of classes, (II) the object color is faded, which leads to wrong classification of the object. and (III) a problem in the light above the camera, causing noise in the object color and thus a wrong classification of the object. All these circumstances lead to performance degradation in the system, and even a failure to achieve the system goal. Hence, we need a solution to maintain the system's resilience and green.

The online learning of the system, the heterogeneous data sources, or the human operator can cast several challenges for both the resiliency and greenery of the system. From the \textit{green} side, the system will require different movements and computational power, leading to more energy consumption and more \coo{} footprint. On the other side, the system performance will be affected when a new unclassified object arrives, which might lead to a late response from the arm and the object will pass the picking area on the conveyor.
The desired solution must trade-off between the \coo{} emission footprints and the time to recover from the underperformance situation.
\enlargethispage{\baselineskip}
\textbf{Proposed Solution}:
The goal of this CAIS, is helping the human to achieve automatic classification of coming objects based on their color with less human interference as possible.
The proposed solution suggests to extends the current online classification workflow by reading a second image data in case the classifier's confidence level about the prediction of the object class is not high, and in this case, there is a need to slow down the conveyor to be able to pick it up within the picking area. If the second image led to a higher confidence level about the second image prediction, the robot will pick the object up and place it in the box, and the human will always have the ability to revert the step and teach the correct classification. Otherwise, if the new image still did not cause a confident prediction, the robot will leave it to the human to classify it but without moving the arm.
This means we can choose between two actions to perform by the robot arm. Recover the system and automatically classify the object, or leave the object for the human to be classified. The two robot arm actions after slowing down the conveyor are ($a_1$) Ask the arm to classify the object, and ($a_2$) Ask the human to classify the object.

\begin{table}[t]
\tiny
\caption{The Gresilience Game Payoffs Matrix}
\begin{center}
\begin{tabular}{|c|c|c|}
\hline
 &\multicolumn{2}{|c|}{\textbf{\textcolor{ForestGreen}{$p_2$}}} \\ \hline
\textbf{\textcolor{red}{$p_1$}} & \textcolor{ForestGreen}{$a_1$} & \textcolor{ForestGreen}{$a_2$} \\ \hline
\cline{2-3}
  &  & \\ 
 \textcolor{red}{$a_1$} & \textcolor{red}{A}, \textcolor{ForestGreen}{b} & \textcolor{red}{C}, \textcolor{ForestGreen}{d} \\ 
  &  & \\ \hline
\cline{2-3}
  &  & \\ 
 \textcolor{red}{$a_2$} & \textcolor{red}{D}, \textcolor{ForestGreen}{c} & \textcolor{red}{B}, \textcolor{ForestGreen}{a}\\ 
  &  & \\ 
  \hline
\end{tabular}
\label{payoffs_matrix}
\end{center}
\end{table}

Both ($a_1$) and ($a_2$) have different approaches to achieve the system goal. The first action is concerned about fast recovery (resilience) by the system and the second action is concerned about energy and minimizing \coo{} footprint (green).
Therefore, the proposed solution is to achieve a green resilient solution, where in case of uncertainty the system will make a decision based on each property preference (player preference), assembling a game theory solution. Game theory has been widely adopted to achieve fast decision-making in finite strict games of online learning \cite{flokas_no-regret_2020, cui_improving_2019}.

Game theory is ``the study of rational decision-making between players who seek the best payoffs for their own interest" \cite{mei_engineering_2017}.
The game players, player resilience ($p_1$) and player green ($p_2$) are taking decisions to maximize their own preferences or in other words their \text{utility functions} \cite{von_neumann_theory_2007}.

The game played by $p_1$ and $p_2$ is similar to the \textit{Battle of Sexes} a coordination game between two players that also has the elements of conflict, and each player's decision affects the other \cite{luce_games_1989}. This proposal introduces a new game named \textit{The Gresilience Game}, in this game, the human and robot arm both have a common goal to classify a set of objects based on their color, and they are collaborating in an online learning environment, where the human teaches the robot how to classify the objects. However, during the learning process, the robot will automatically classify objects, that it already learned, but in case of uncertainty, the robot faces some challenges in reaching a high \textit{confidence level} ($\epsilon$ hereinafter, where $\epsilon \in [0, 1]$) of classification about the object's color and deciding whether it should leave the classification to the human, or react and recover. The system in this situation is facing a dilemma because it wants to maintain its performance and move the arm to classify the object, which maximizes the system's resilience. However, if it does and makes a mistake this means wasting more time as the human needs to validate and correct that movement, and also more \coo{} footprint from the extra energy of the wrong movement.

The system founds itself in a situation of choosing between being resilient and having less human interaction, where both players go with $a_1$, or being green by choosing $a_2$ and leaving the human to do the job. However, the situation is not that simple, because choosing $a_1$ by both players means that we can trust the classifier's decisions and there will be no need for the human to correct it, or in other words, the value of $\epsilon$ is high ($\epsilon \approx 1$). On the other hand, when the value of $\epsilon$ is low ($\epsilon \approx 0$), then we cannot trust the classifier's decisions and it is better to ask the human to do the job, and thus choose $a_2$ by both players. But when the value of $\epsilon$ is almost in the middle ($\epsilon \approx 0.5$), then choosing between $a_1$ and $a_2$ becomes a not straightforward move. \textit{The Gresilience Game} is a probability distribution game that has two pure strategy Nash equilibria (PSNE), where both players choose the same action, and a mixed strategy Nash equilibrium (MSNE) based on the probability of each player's action \cite{von_neumann_theory_2007, stowe_cheating_2010}. 
\enlargethispage{\baselineskip}

\textit{The Gresilience Game} payoffs are computed based on each player's preference and the selected action. Table~\ref{payoffs_matrix} shows a general form of the game payoffs matrix, where $A > B > C > D$ and small letters are used for readability. The matrix shows two PSNE for choosing $a_1$ by both players and choosing $a_2$ by both as well. $A$ means the player has chosen his best action, which allows him to utilize all the system factors to achieve resilience and green. In contrast, $B$ means that although the player did not choose his best action, which means he will not be able to utilize all the system factors, he is still happy choosing a similar action as the other player. On the other hand, $C$ and $D$ mean different actions chosen by the players, which means worse utilization of the system factors. The system factors are the needed time to classify the object by the human ($t_h$), the needed time to classify it by the arm ($t_a$), the reduction of human interaction ($h$), and the reduction of \coo{} footprint (\coo{}). Equations Eq.~\eqref{eq_a}  - \eqref{eq_d} show how to compute each payoff in regard to $\epsilon$, noting that the sign of a system factor simply indicates whether it is a good or bad utilization by the chosen action for that player.
\begin{equation}
A = \epsilon * (t_h + t_a + h + \coo{})
\label{eq_a}
\end{equation}
\begin{equation}
B = \epsilon * (t_h + t_a + \coo{} - h)
\label{eq_b}
\end{equation}
\begin{equation}
C = \epsilon * (t_h + \coo{} - h)
\label{eq_c}
\end{equation}
\begin{equation}
D = \epsilon * (t_h + \coo{} - t_a - h)
\label{eq_d}
\end{equation}
Now lets consider the MSNE of \textit{The Gresilience Game}, where we need to find the expected utility ($EU$) for $p_2$ of choosing $a_1$ denoted ($EU{p_2.a_1}$) and what is the $EU$ for $p_2$ to choose $a_2$ denoted ($EU{p_2.a_2}$). Then by setting these two expected utilities to equal each other, we can calculate the MSNE for $p_1$. Eq.~\eqref{eu_p_2_a_1} shows that $EU{p_2.a_1}$ depends on the probability ($\sigma$) of $p_1$ choosing $a_1$ times the payoff $p_2$ gains when choosing $a_1$ ($c$), added to the probability of $p_1$ choosing $a_2$ ($1 - \sigma$) times the payoff $p_2$ gains when $p_1$ choosing $a_2$. The same steps are followed to find $EU{p_2.a_2}$ in Eq.~\eqref{eu_p_2_a_2}.
\begin{equation}
EU{p_2.a_1} = \sigma_{p_1.a_1} * b + (1 - \sigma_{p_1.a_1}) * c
\label{eu_p_2_a_1}
\end{equation}
\begin{equation}
EU{p_2.a_2} = \sigma_{p_1.a_1} * d + (1 - \sigma_{p_1.a_1}) * a
\label{eu_p_2_a_2}
\end{equation}
Next, to find the MSNE for $p_1$, let Eq.~\eqref{eu_p_2_a_1} equals Eq.~\eqref{eu_p_2_a_2}, and by solving the formula, the value of $\sigma$ is shown in Eq.~\eqref{sig_p_1}, where $a + b - c - d \neq 0$ and positive since $a - c > 0$ and $b - d > 0$ because $a > c$ and $b > d$ and so the summation is also positive, and $a - c$ is positive since $a > c$ , and finally $a + b - c - d > a - c$ because its the same as saying $b - d > 0$ by adding $c$ and subtracting $a$ from both sides, and it is the same as $b > d$ by adding $d$ to both sides that always holds, which means we have a valid MSNE probability for $p_1$, and thus, $\sigma_{p_1.a_1} \in [0, 1]$.
\begin{equation}
\sigma_{p_1.a_1} = (a - c) / (a + b - c - d)
\label{sig_p_1}
\end{equation}
\begin{equation}
\sigma_{p_2.a_1} = (B - C) / (A + B - C - D)
\label{sig_p_2}
\end{equation}
By following the same steps to find the MSNE for $p_2$, Eq.~\eqref{sig_p_2} shows the MSNE probability equation of $p_2$ choosing $a_1$, where $\sigma_{p_2.a_1} \in [0, 1]$. Hence the payoff for $p_1$ in a MSNE is shown in Eq.~\eqref{p_1_mix}, and $p_2$ payoff is shown in Eq.~\eqref{p_2_mix}.
\begin{equation}
\begin{split}
p_{1_{msne}} & = (A + B - C - D) * (\sigma_{p_1.a_1} * \sigma_{p_2.a_1}) \\
& + (C - B) * \sigma_{p_1.a_1} + (D - B) * \sigma_{p_2.a_1} 
+ B
\end{split}
\label{p_1_mix}
\end{equation}
\begin{equation}
\begin{split}
p_{2_{msne}} & = (a + b + c + d) * (\sigma_{p_1.a_1} * \sigma_{p_2.a_1}) \\
& + (d - a) * \sigma_{p_1.a_1} + (c - a) * \sigma_{p_2.a_1} 
+ a
\end{split}
\label{p_2_mix}
\end{equation}

The MSNE payoffs in both Eq.~\eqref{p_1_mix} and Eq.~\eqref{p_2_mix} shows that even when the decision is not clear, the system will still maintain its resilience and green when the system randomly chooses based on the probability of the system a mix between $a_1$ and $a_2$, which leads to maximizing the benefits from the chance that robot arm may correctly classify the object and reducing the time to recover and \coo{} footprint.
\enlargethispage{\baselineskip}
\section{Conclusion and Future Work}
\textbf{Conclusion}: This doctorate proposal suggests a solution using game theory to solve the trad-off between maximizing resilience and recovering the system after a performance degradation and maximizing green by minimizing the system \coo{} footprint, by considering both resilience and green adversaries.

The proposed solution is reflected over a special type of cyber-physical system, which is a collaborative artificial intelligence system that involved both robots and humans, and thanks to the cooperation toward a unique goal it made the use of game theory more suitable.

\textbf{Future Work}: After implementing the proposed solution, we will explore new methods based on optimization to address the trade-off between \textit{green} and \textit{resilience}, and compare the new model with the solution suggested by this proposal.
Additionally, more work will be done to extend the case study to be able to classify the objects based on their shape, which requires running other learners to predict the object shape. This extension adds more complexity to online learning resilience and green. In this context a discussion about how we can utilize server-less cloud functionality to initiate what we call \textit{on-demand learning} to save energy consumption and reduce \coo{} footprint.

\begin{figure}[b]
\centerline{\includegraphics[width = 0.35 \textwidth,trim={0 14 1 1},clip]{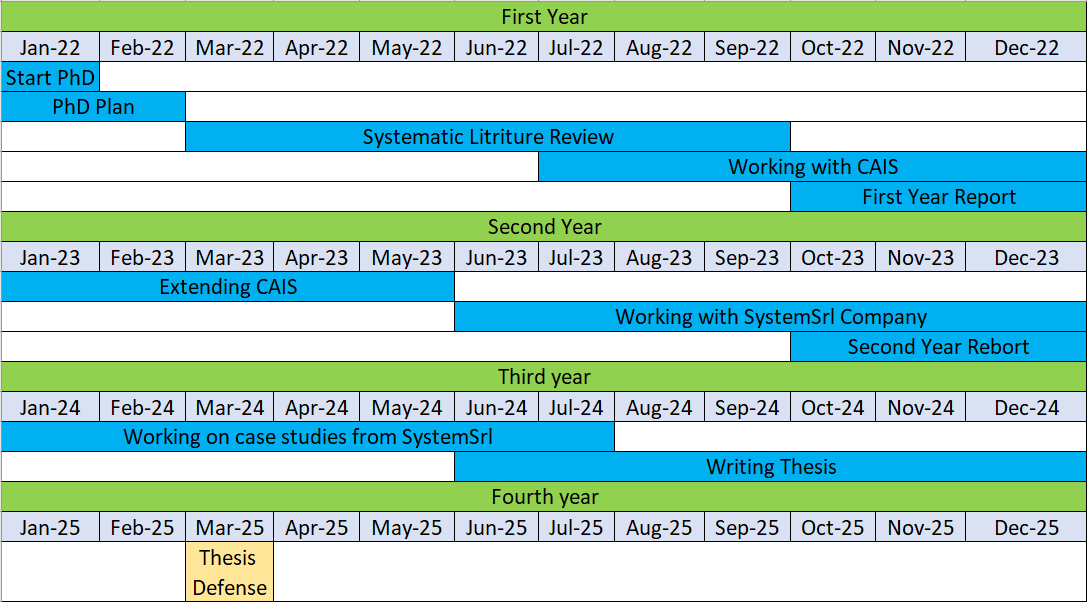}}
\caption{PhD timeline}
\label{timeline}
\end{figure}

\textbf{Timeline}: This PhD kicked off in January 2022, and at the time this proposal was created the work is still in the literature review stage. Fig.~\ref{timeline} shows the timeline of the PhD, where the expected date of defense is March 2025.
\section*{Acknowledgment}
This research is supervised by Russo Barbara, Full Professor, and Robbes Romain, Associate Professor at the Faculty of Computer Science of the Free University of Bozen-Bolzano, Italy. For the study of PhD in Advanced-Systems Engineering.

\printbibliography

\end{document}